\documentclass[twocolumn,preprintnumbers,amsmath,amsfonts,amssymb,notitlepage,showpacs,aps,prb,longbibliography]{revtex4-2}

\usepackage{color}

\usepackage{amsmath}

\usepackage{graphicx}
\usepackage[monochrome]{xcolor}

\usepackage{hyperref}
\usepackage{mathtools}

\usepackage[normalem]{ulem}

\def \be{\begin{equation}}
\def \ee{\end{equation}}
\def \ba{\begin{array}}
\def \ea{\end{array}}
\def \bea{\begin{eqnarray}}
\def \eea{\end{eqnarray}}

\date{\today}
\begin{document}
\title{
Fractal structure of multipartite 
entanglement in monitored quantum circuits}

\author{Vaibhav Sharma}
\email{vaibhavsharma@rice.edu}
\affiliation{Smalley-Curl Institute, Rice University, Houston, TX 77005 and Department of Physics and Astronomy, Rice University, Houston, TX 77005}
\author{Erich J Mueller}
\email{em256@cornell.edu}
\affiliation{Laboratory of Atomic and Solid State Physics, Cornell University, Ithaca, NY 14853}

\begin{abstract}

We study the structure of multipartite entanglement in monitored quantum circuits exhibiting measurement-induced phase transitions (MIPTs). Using a one-dimensional Clifford circuit subject to local measurements with a probability $p$, we show numerically that the entanglement depth, corresponding to the size of the largest cluster of entangled qubits scales as a power law with system size on both sides of the transition. The power law exponent is 1 in the entangling phase and continuously decreases to 0 as $p \to 1$ in the disentangling phase. In addition, we find that the spatial support of the largest cluster exhibits an approximate fractal geometry with a tunable fractal dimension controlled by the measurement rate. We argue that this structure arises from a competition between unitary-driven coagulation of entangled clusters and measurement-induced fragmentation, giving rise to a fractal steady state reminiscent of classical coagulation–fragmentation models. Away from the MIPT critical point, the fractal dimension matches the entanglement depth power law exponent.  These results show that multipartite entanglement structure provides a fresh perspective on the emergent quantum correlations in monitored quantum circuits and noisy quantum dynamics. 

\end{abstract}

\maketitle

\section{Introduction}

Monitored quantum circuits with random unitary gates and projective measurements have emerged as powerful tools to understand dynamics of entanglement in noisy quantum many-body systems~\cite{randomcircuits}. By varying the parameters of these monitored circuits, such as the relative probabilities of unitary operations and projective measurements, these circuits can exhibit a variety of many body phases and quantum phase transitions between them~\cite{trans1,trans2,trans3,trans4,measonly1,measonly2,measonly3,measonly4,measonly5,measonly6,measonly7,measonly8}. An iconic example is the measurement-induced entanglement phase transition from a volume law entangled phase to an area law entangled phase~\cite{trans1,trans2,trans3,trans4}. These entanglement phase transitions are typically characterized using bipartite diagnostics such as entanglement entropy, which distinguish phases through their scaling with system size. Here we instead study the spatial distribution and geometry of multipartite entanglement in such circuits, finding a fractal structure.

While bipartite entanglement entropy can distinguish states based on the amount of entanglement across a partition, it does not capture the spatial range of entanglement or the structure of multipartite correlations in the state. As an example, consider the $N$-qubit Greenberger-Horne-Zeilinger (GHZ) state on a one-dimensional chain, $\Psi_{GHZ}=|111\cdots\rangle+|000\cdots\rangle$. It is an area law state as the bipartite entanglement entropy across any partition is independent of the system size $N$. Such area law behavior indicates that the state contains a very small amount of entanglement. Nonetheless the GHZ state is an example of genuine $N$-partite entanglement; every qubit is entangled with every other qubit. The bipartite entanglement entropy is insensitive to this feature, and cannot by itself be used to characterize multipartite entanglement. This motivates characterizing the entanglement properties of monitored circuits beyond bipartite entanglement entropy measures. 

Some previous works have examined multipartite entanglement in systems exhibiting measurement induced phase transitions using quantum Fisher information~\cite{qfimonitored,qfimonitored1,qfimonitored2,qfimonitored3}, tripartite entanglement measures~\cite{tripartite,tripartite1} and $k$-party mutual information~\cite{kparty}. Quantum Fisher information provides a lower bound to the number of parties entangled in a state. While it is a valuable tool, it is strongly affected by the choice of the operator used to compute it~\cite{qfi}. Choosing local operators can often lead to an underestimation of these bounds~\cite{nonlocalqfi}. Tripartite entanglement was found to show some genuine three-qubit long-range entanglement near the measurement induced phase transition~\cite{tripartite} but it does not easily generalize to $N$-partite entanglement computations. These approaches also do not directly characterize the geometry of multipartite entanglement. Here we explore multipartite entanglement structure by calculating the entanglement depth, which corresponds to the size of the largest cluster of entangled qubits. \textcolor{blue}{The entanglement depth essentially quantifies the minimum circuit depth needed for a quantum circuit with spatially local unitary gates to produce this largest cluster. This quantity} is basis agnostic and does not require choosing a particular operator or system partition. 

We study circuits where random two-qubit Clifford unitaries are  interspersed with single-qubit Pauli measurements. In a given circuit layer, the latter occur with a probability $p$. This circuit exhibits a measurement-induced entanglement phase transition between the area and volume law phase~\cite{trans2,trans3}. By using the formalism developed in~\cite{entstructures}, we extract the largest non-separable cluster of qubits. The size of that cluster is known as the entanglement depth. We find that the entanglement depth scales as a power law with system size in both the area and the volume law phase, indicating persistence of long range entanglement even within the area law phase. Moreover, we find that the spatial support of the largest qubit cluster exhibits an approximate fractal structure whose fractal dimension tracks the entanglement depth power law exponent away from the entanglement phase transition critical point. This hints towards a self similar organization of multipartite entanglement arising from a competition between unitary gate driven coagulation and measurement-induced fragmentation of entangled clusters.

Our results highlight multipartite entanglement structure as a complementary lens for understanding monitored quantum dynamics. Whereas prior work characterized measurement-induced transitions largely through bipartite entropy scaling, our analysis shows that many-body correlations can organize into self-similar, fractal structures. This perspective positions entanglement depth and the geometry of entangled clusters as informative diagnostics of quantum correlation structure under measurements and noise.

\section{Circuit Model}

Figure~\ref{fig:entstruc}(b) shows our circuit model where a one-dimensional chain of $L$ qubits evolves under random two-qubit Clifford unitaries and single-site projective measurements in a brickwork arrangement~\cite{trans2,trans3,trans4}. At each odd (even) time step, nearest-neighbor qubits connected by odd (even) bonds are acted upon by a random two-qubit Clifford unitary. Then at each step, every qubit is projectively measured in the $z$-basis with a probability $p$. Due to the Gottesman-Knill theorem~\cite{gottknill}, the states generated by this circuit are stabilizer states which can be efficiently simulated by classical computers despite extensive entanglement. For system size $L$, the stabilizer formalism keeps track of the quantum state by storing $L$ binary strings of length $2L$~\cite{measonly8}. We start with a product state in the $z$-basis and evolve under the random Clifford circuit at a given measurement probability $p$.  We proceed for $4L$ steps to reach a steady state. Our results are obtained by ensemble averaging over 500 such steady state realizations for system sizes up to $L = 240$.  

\section{Entanglement Structure Calculation}

We use the method developed in~\cite{entstructures} to extract the entanglement depth and construct the largest cluster of entangled qubits.  Our approach involves constructing an {\em entanglement structure diagram} which gives a hierarchical representation of how various qubits are entangled with one-another.  It contains information that we do not use here, but nonetheless is an efficient method of constructing the largest cluster of entangled qubits. Here, we briefly recap the strategy.  It involves grouping qubits into clusters characterized by a parameter $w$. Such a $w$-cluster has the property that each qubit within it has non-zero {\it total correlations} with at least $(w-1)$ other qubits in the cluster; the total correlation of $w$ qubits is $I = \sum_{i =1}^{w} S_i - S_{\rm cluster}$. Here $S_i$ is the entanglement entropy of qubit $i$ with respect to the rest of the system and $S_{\rm cluster}$ is the entanglement entropy of the $w$ qubit group with the rest of the system. 

This construction is iterative where a cluster found in one iteration becomes an indivisible new element in the subsequent iterations. We can then continue to recursively group these clusters into new clusters until all qubits have been assigned to large decoupled clusters. The overall state can be written as a tensor product state over these decoupled clusters. Each large cluster itself describes a set of qubits that are genuinely multipartite entangled. The size of the largest cluster denotes the entanglement depth of the state. 

\begin{figure}
    \includegraphics[width=\columnwidth]{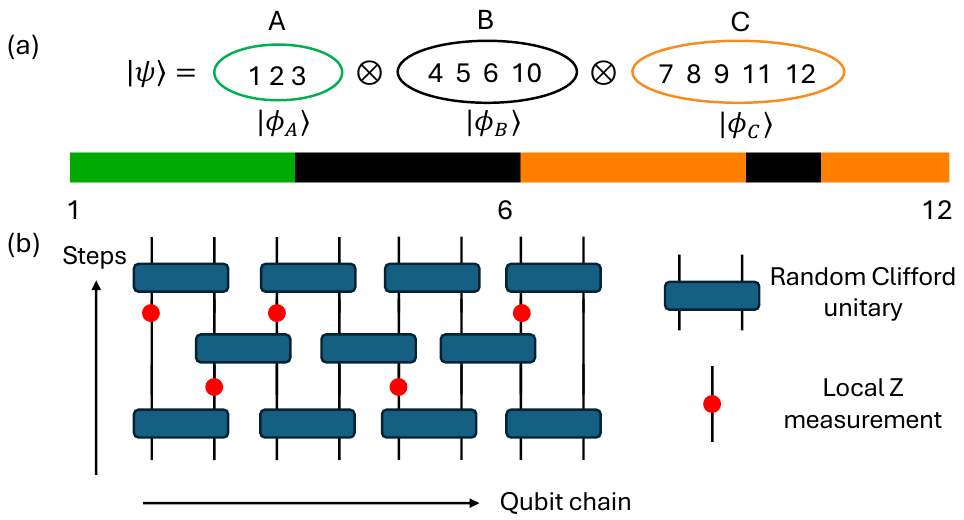}
    \caption{(a) Entanglement structure schematic of a twelve-qubit state $|\psi\rangle$. The state is separable into a tensor product over clusters A (green),B (black) and C (orange) denoted as $|\psi\rangle = |\phi_A\rangle \otimes |\phi_B\rangle \otimes |\phi_C\rangle$. The colored bar shows the qubits arranged in their native spatial order. The entanglement depth of this state is 5, corresponding to the size of the largest cluster C. (b) Monitored circuit in brickwork form with two-qubit random Clifford unitaries and local projective $z$-basis measuremements with probability $p$.}
    \label{fig:entstruc}
\end{figure}

Fig.~\ref{fig:entstruc} shows an abreviated schematic of an entanglement structure of some twelve qubit state, $|\psi\rangle$. There are three large decoupled clusters over regions $A$, $B$ and $C$ suggesting that $|\psi\rangle = |\phi_A\rangle \otimes|\phi_B\rangle \otimes |\phi_C\rangle$. Here $|\phi_A\rangle$, $|\phi_B\rangle$ and $|\phi_C\rangle$ denote some genuine multipartite entangled states that cannot be further written as tensor product states over smaller subsystems. The entanglement depth of the state in this schematic example is $5$ as it corresponds to the size of the largest cluster, denoted by region C. 

As a function of system size, this method scales exponentially for unitary-only circuits and polynomially for local measurement-only circuits~\cite{entstructures}. The computational time can be further improved by coarse graining the system. For that, instead of considering each qubit individually in the first iteration of the method, we can group a few neighboring qubits together, and then consider these groups as the individual entities in the clustering method. 

\section{Entanglement depth and long-range entanglement}

We calculate the average entanglement depth of the steady state ensemble generated by the monitored Clifford circuit as a function of the single-site measurement probability $p$ and system size $L$. A growth in entanglement depth with system size  indicates the presence of long-range entanglement. In our analysis, we use a two-qubit coarse graining procedure before evaluating the entanglement structure diagram and extracting the entanglement depth. There are two reasons for using this coarse graining. First, due to the two-qubit unitary gates in our circuit, there are always neighboring pairs of qubits which are entangled, giving a lower bound of $2$ on the entanglement depth in the absence of coarse graining.  A system size dependent population of such neighboring pairs could erroneously be mistaken for a sign of long-range entanglement. Second, coarse graining over pairs of sites leads to shorter computational times and lets us go to higher system sizes when $p$ is sufficiently close to the volume law phase. 

\begin{figure}
    \includegraphics[width=\columnwidth]{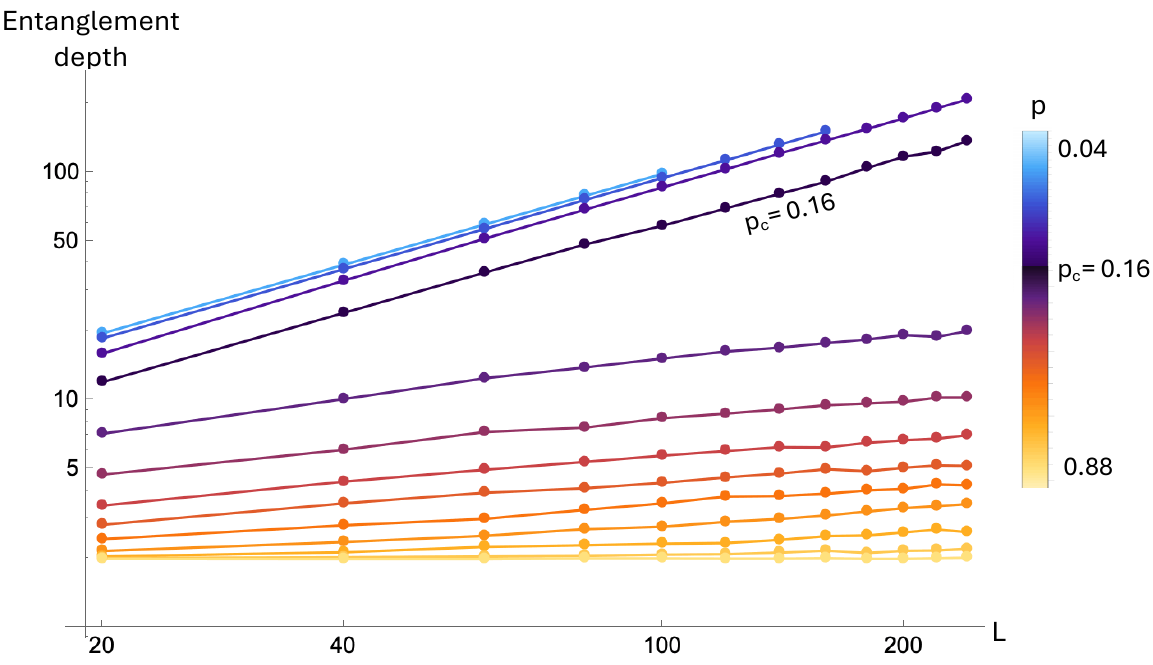}
    \caption{Log-log plot of average entanglement depth as a function of system size $L$ for the steady state ensemble of states formed by applying random nearest-neighbor two-site clifford gates, interspersed with random single site measurements.  The probability of a local measurement is $p$. The curves are plotted in steps of $0.04$ for $p < p_c$ and in steps of $0.08$ for $p > p_c$.  There is a phase transition at $p=p_c\sim 0.16$ \cite{trans2,trans3}.
    }
    \label{fig:depthcurves}
\end{figure}

Figure~\ref{fig:depthcurves} shows the entanglement depth as a function of system size, $L$ for various measurement probabilities, $p$ on a log-log plot. We can see that the entanglement depth grows as a power law both within the volume law phase ($p < p_c = 0.16$) and the area law phase where $p > p_c$. This scaling captures the presence of long-range entanglement within the area law phase that is not manifest in bipartite entanglement measures. The slope on the log-log plot seems to be nearly constant when $p \leq p_c$ and reduces as $p$ increases within the area law phase. We note that deep within the volume law phase, we limit ourselves to small $L$ due to the computational difficulty of calculating the entanglement structure for volume law states. 

\begin{figure}
    \includegraphics[width=\columnwidth]{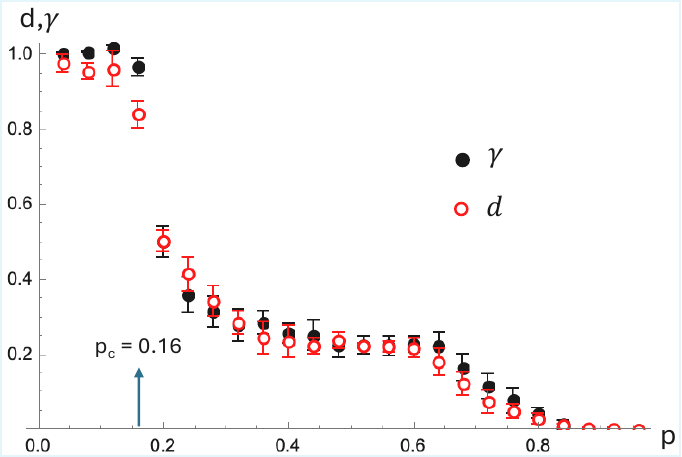}
    \caption{Power law exponent of entanglement depth growth with system size, $\gamma$ \textcolor{blue}{(shown as black dots)} and fractal dimension, $d$ \textcolor{blue}{(shown as open red circles)} of the largest cluster of entangled qubits as a function of measurement probability, $p$. The critical point for the volume law-area law phase transition is shown as $p_c = 0.16$.}
    \label{fig:exponentdim}
\end{figure}

We use a robust linear regression technique called the Thiel-Sen estimator~\cite{theil,sen} to fit this data to a straight line and extract the power law exponent $\gamma$ such that $D\propto L^\gamma$, where $D$ denotes the entanglement depth. In Appendix~\ref{app:thielsen}, we discuss the Thiel-Sen estimator for regression and how we get error bars for the extracted slopes. The results are shown as blue data points in Fig.~\ref{fig:exponentdim}. We can see that the power law exponent is $\gamma=1$ within the volume law phase and at the critical point. In the area law phase, $\gamma$ continuously decreases to $0$ as $p \to 1$. The plot shows a sharp drop in the power law exponent at the volume law - area law transition at $p_c = 0.16$. For $0.3 \lesssim p \lesssim 0.6$, there appears to be a plateau in $\gamma$.  There is a knee at $p\sim0.6$. For $p \gtrsim 0.6$, the slopes decrease to $0$ as we approach $p = 1$ where all qubits are expected to be disentangled with each other. 

The average entanglement depth denotes the ensemble average of the largest cluster size in each realization of the random circuit. In the volume law phase, there is extensive entanglement and the bipartite entanglement entropy itself grows linearly with system size. Thus we expect the largest cluster to also grow linearly with system size, consistent with our observation that $\gamma=1$ for $p<p_c$. At $p=0$, we expect all qubits to be entangled and the largest cluster exactly spans the entire system. Beyond the volume law phase, $p>p_c$, the bipartite entanglement entropy either grows logarithmically with system size (at critical point) or saturates quickly to a constant. However, we find that the largest cluster continues to grow with system size even in these regimes.

The number of entangled spins in the largest cluster grows, despite the saturation in the number of clusters which cross any bipartition.  This suggests that there are a small number of very large clusters.  Increasing the frequency of measurements, by increasing $p$, shrinks the size of the largest cluster, but it still grows with $L$.  There is a marked robustness to these states, which is very different from what we saw in the GHZ example.  A single measurement on a GHZ state results in a disentangled state. 

As noted, Fig.~\ref{fig:exponentdim} shows a distinct knee near $p=0.6$.  This occurs when the ensemble averaged entanglement depth is of order, $D = 2$.  For larger $p$, the ensemble predominantly contains realizations where $D=2$.  At smaller $p$, one instead has a broader distribution of $D$.  The location of the knee is sensitive to the degree of course graining that we perform.

\section{Fractal Structures}

Here we study the spatial structure of the largest entangled cluster. Earlier we showed that the size of this cluster $D$ grows with system size as $D\sim L^\gamma$, with $0<\gamma<1$ in the area law phase.  This power law is suggestive of a fractal shape, analogous to a Cantor set.

\begin{figure}
    \includegraphics[width=\columnwidth]{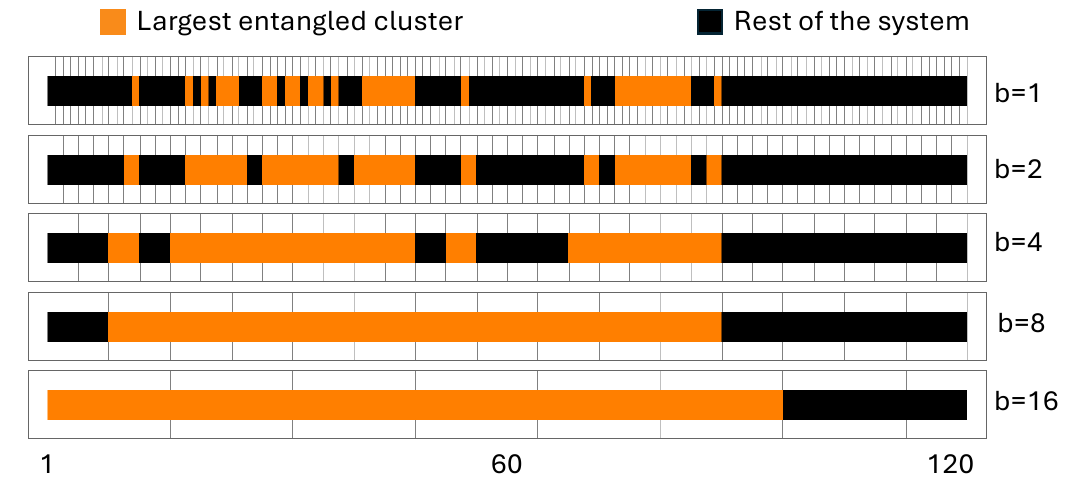}
    \caption{
    A typical 120 qubit state generated at measurement probability, $p=0.2$. The same state is shown with different coarse-graining scales, denoted by gridlines enclosing $b$ qubits. For a given coarse-grained picture, the orange squares denote qubits that are part of the largest entangled cluster while black squares denote the remaining qubits. The orange cluster spans the entire system, but is filled with 'holes' of multiple sizes that break it into discontinuous pieces.
    }
    \label{fig:fractals}
\end{figure}

Figure~\ref{fig:fractals} shows 
a typical 120-qubit state generated by our monitored circuit model with the measurement probability, $p=0.2$. The qubits are arranged in their native spatial order.
The orange squares denote qubits which are part of the largest entangled cluster while the black squares denote the rest of the qubits. Without coarse-graining ($b=1$), we can see that the largest cluster is not a continuous block but instead contains a number of unentangled qubits which appear as holes in the cluster. These holes span multiple scales, as is illustrated by the coarse-grained clusters characterized by different values of $b$ as shown in Fig.~\ref{fig:fractals}.  The presence of holes at multiple coarse grained levels implies that the largest cluster of entangled qubits is a fractal object. The observed pattern of holes is analogous to ones seen in mathematical fractals such as the Sierpienski triangle where there are triangle-shaped holes at every coarse-grained level. An even better analogy could be made with random Cantor sets \cite{cantor}.  Regardless, visually the pattern in Fig.~\ref{fig:fractals} is suggestive of a fractal.

We use a box counting method to extract the fractal dimension of the largest cluster. For a given state in our ensemble, we first coarse-grain by grouping $b$ neighboring qubits into indivisible boxes. We then extract the entanglement structure diagram to get the coarse-grained entanglement depth.
The detailed procedure is explained in Appendix~\ref{app:boxcounting}. We count the number of boxes, $N_b$ that are required to tile the largest entangled cluster (such as the orange region in Fig.~\ref{fig:fractals}). The fractal dimension, $d$ is then defined as, $d = -\ln{N_b}/\ln{b}$. 

\begin{figure}
    \includegraphics[width=\columnwidth]{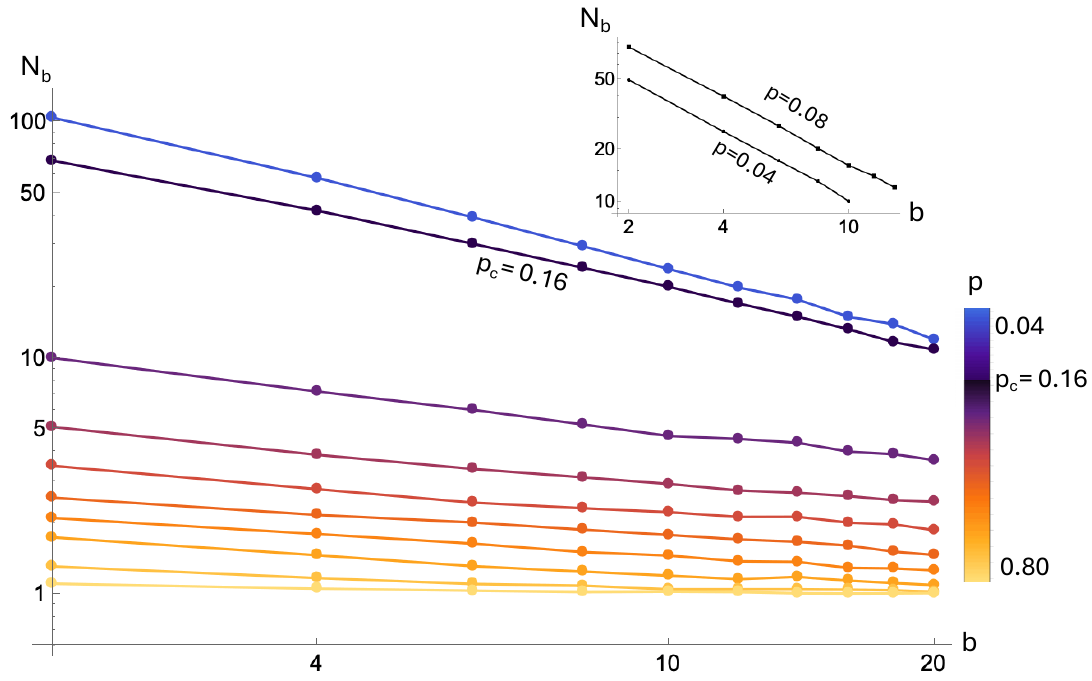}
    \caption{Log-log plot of number of boxes, $N_b$ as a function of box size, $b$ for various measurement probabilities, $p$. The curves are plotted in steps of $0.04$ for $p < p_c$ and in steps of $0.08$ for $p > p_c$. The inset separately shows the same data deep in the volume law phase for $p=0.04,p=0.08$.}
    \label{fig:boxcountingcurves}
\end{figure}

We use box sizes ranging from $2$ to $20$ and calculate the ensemble-averaged value $N_b$ for different measurement probabilities $p$. We used a system size $L = 240$ for $p \geq 0.12$ and the results are shown in a log-log plot in Fig.~\ref{fig:boxcountingcurves}. We can see that the log-log plots show a linear relationship between $ \ln N_b$ and $\ln b$ with a negative slope, suggesting a fractal structure. \textcolor{blue}{We note that the apparent gap in Fig.~\ref{fig:boxcountingcurves} below the $p_c=0.16$ line is simply due to the coarseness in the values of $p$ used in this figure.}

The magnitude of the slope of the curve gives the fractal dimension. We again used the Thiel-Sen estimator to extract the fractal dimension from these data. Deep within the volume law phase, we use smaller system sizes ($L = 100,160$ for $p=0.04,0.08$ respectively) to extract the fractal dimension. 

\begin{figure}
    \includegraphics[width=\columnwidth]{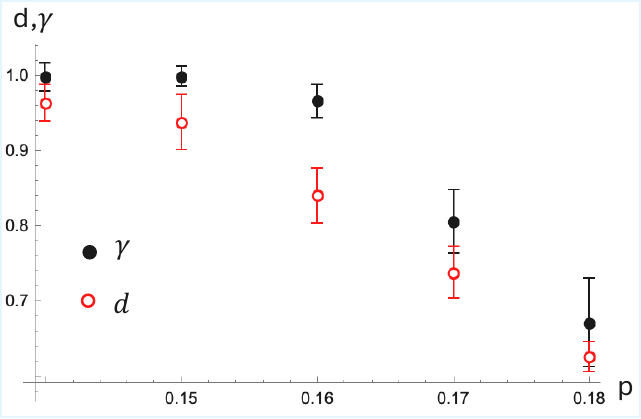}
    \caption{Power law exponent of entanglement depth growth with system size, $\gamma$ \textcolor{blue}{(shown as black dots)} and fractal dimension, $d$ \textcolor{blue}{(shown as open red circles)} of the largest cluster of entangled qubits as a function of measurement probability, $p$ near the entanglement phase transition critical point, $p_c = 0.16$.}
    \label{fig:critexponentdim}
\end{figure}

The results are shown as yellow data points in Fig.~\ref{fig:exponentdim}. We can see that the largest entangled clusters are fractal objects with a fractal dimension between 0 and 1. In the area law phase, the fractal dimension agrees with the entanglement depth exponent up to the error bars which are inferred from the accuracy of the power law fit. At the critical point ($p = 0.16$), there is however a  small, but statistically significant, difference between the two.  Fig.~\ref{fig:critexponentdim} shows the region close to the critical point and we see that the fractal dimension and entanglement depth power law exponent extracted from our fits do not agree close to the critical point. This discrepancy reduces as we go away from the critical point deeper into the volume law phase.  These discrepancies may be related to logarithmic corrections to scaling, finite size effects, or some other additional structure which emerges at the critical point. \textcolor{blue}{An interesting future study could be to analyze the critical point using other entanglement structure diagnostics such as $k$-uniformity and entanglement structure layers~\cite{entstructures}.}

In our circuits, the fractal structure emerges from a competition between unitary gate-driven cluster growth and measurement-induced fragmentation. A unitary acting between a qubit in the largest cluster and one outside typically merges clusters, while a local measurement can remove a qubit or, in extreme cases, dissolve an entire cluster, as exemplified by the GHZ state. Thus it is non-trivial to deduce the amount by which the cluster grows/shrinks. The interplay of these growth and fragmentation processes produces a steady state in which qubits are entangled hierarchically, where there are structures on multiple scales that give rise to approximate fractal geometries. Such structures are reminiscent of classical coagulation–fragmentation models in statistical physics, where competing aggregation and break-up dynamics produce steady, self-preserving cluster size distributions with scale-invariant distributions and fractals~\cite{fragment1,fragment2}. \textcolor{blue}{As shown in Figs.~\ref{fig:exponentdim},\ref{fig:critexponentdim},} the fractal dimension of the largest entangled cluster can be tuned continuously by the measurement probability 
$p$, reflecting the balance between unitary-driven coagulation and measurement-induced fragmentation.

We emphasize that the fractals we see here are \textit{approximate} physical fractals and not exact mathematical fractals. An exact mathematical fractal would show the same structure at arbitrarily small and large scales. In our case, the smallest physical scale we have is a single qubit. Our finite system size limits the largest possible scale. Thus, for system size $L = 240$, our box sizes can cover only a few powers of $2$ ranging from $b=2$ to $b=20$. In this intermediate size regime, the ensemble averaged largest cluster of entangled qubits is well described as a fractal object. While we believe that this behavior will extend to the thermodynamic limit, it is possible that we are observing a finite size effect which would not be present in numerical experiments on larger systems. \textcolor{blue}{For the moderately large system sizes we have studied with open boundary conditions, we do not expect any significant edge effects in our results.}

\section{Summary and Outlook} 

Given that monitored circuits exhibit entanglement phase transitions, it is natural to ask about the resulting structure of multipartite entanglement. 
In this work, we use the method developed in~\cite{entstructures} to calculate the entanglement depth of the ensemble of states generated by random Clifford circuits where random two-qubit unitary gates are interspersed with single qubit projective measurements with probability $p$. 

We find that significant multipartite entanglement survives even within the area law phase. Across both the volume and area law phases, we find that the entanglement depth scales as a power law with system size, demonstrating that long-range entanglement and multipartite entanglement are more robust to measurements than bipartite entanglement entropy. Furthermore, the largest entangled qubit cluster exhibits an approximate fractal geometry, with a fractal dimension that coincides with the entanglement depth exponent away from the critical point. This reflects the competition between unitary-driven cluster growth and measurement-induced fragmentation, producing hierarchical, self-similar structures reminiscent of classical coagulation–fragmentation models.

This geometric perspective on multipartite entanglement provides a complementary lens for analyzing quantum correlations in monitored dynamics, beyond traditional bipartite diagnostics such as entanglement entropy or mutual information. It highlights how complex, scale-invariant structures can emerge naturally in noisy quantum systems, with potential relevance for understanding quantum correlations in hybrid quantum circuits, noisy quantum devices, and open quantum systems. Extending this multipartite entanglement approach to higher-dimensional circuits and non-stabilizer states represents a promising avenue for future exploration. 

\section{Acknowledgements}

VS acknowledges support from the J. Evans Attwell Welch fellowship by the Rice Smalley-Curl Institute. EJM acknowledges support from the National Science Foundation under Grant No. PHY-2409403.

\appendix

\section{Thiel-Sen Linear Regression}\label{app:thielsen}

Here we briefly describe the Thiel-Sen regression method we used to extract the slope and the error bars corresponding to the straight line fits of the entanglement depth and box-counting data in the log scale in Fig.~\ref{fig:depthcurves} and Fig.~\ref{fig:boxcountingcurves}. Given a set of data points $\{y_i,x_i\}$, we first calculate the slopes $m_{i,j}=(y_j-y_i)/(x_j-x_i)$ obtained from all possible pairs of data points. Only pairs with distinct values of $x_i$ and $x_j$ are considered. Here $i$ and $j$ are indices identifying the data points.

The median, $m$ of the data set $m_{i,j}$ is the Thiel-Sen estimator of the slope for the straight line fit. The error in the estimated slope is given by the median absolute deviation of this data set. It is obtained by calculating the median of the data set, $\{|m-m_{i,j}|\}$. This quantity is also a measure of statistical dispersion, analogous to the standard deviation that averages the squared distance from the mean of a data set. For normal distributions, the median absolute deviation and standard deviation are proportional~\cite{mad}. 

In contrast to the least-squares method, the Thiel-Sen method is insensitive to outliers and works well for noisy data that is not normally distributed, making it a more robust linear regression method~\cite{thielsenadv}.  For the data here, however, the two approaches give similar results.

\section{Box Counting Method}\label{app:boxcounting}

Here we describe the box counting method we used to extract the fractal dimension of the largest entangled cluster. For a given state generated by our random circuit, we first coarse-grain our system by grouping $b$ neighboring qubits together. With qubits labeled by numbers corresponding to their native spatial order, we group qubits labeled from 1 to $b$, then from $b+1$ to $2b$ and so on for a system size $L$, giving us $\lceil L/b \rceil$ boxes. These boxes are considered as indivisible entities, coarse-graining any entanglement structure that we obtain. 

We then compute the coarse-grained entanglement structure using the method in ~\cite{entstructures}. The resulting structure for different coarse-graining scales $b$ shown in Fig.~\ref{fig:fractals} where the gridlines show the box sizes visually. We count the number of boxes involved in the largest cluster within this entanglement structure and term it as $N_b$. We repeat the calculation with different box sizes $b$, extracting $N_b$ as a  function of $b$ from each of the entanglement structures. 

\nocite{dataset}
\bibliography{main}

@article{randomcircuits,
   author = "Fisher, Matthew P.A. and Khemani, Vedika and Nahum, Adam and Vijay, Sagar",
   title = "Random Quantum Circuits", 
   journal= "Annual Review of Condensed Matter Physics",
   year = "2023",
   volume = "14",
   number = "Volume 14, 2023",
   pages = "335-379",
   doi = "https://doi.org/10.1146/annurev-conmatphys-031720-030658",
   url = "https://www.annualreviews.org/content/journals/10.1146/annurev-conmatphys-031720-030658",
   publisher = "Annual Reviews",
   issn = "1947-5462",
   type = "Journal Article",
  }

@article{trans1,
  title = {Quantum Error Correction in Scrambling Dynamics and Measurement-Induced Phase Transition},
  author = {Choi, Soonwon and Bao, Yimu and Qi, Xiao-Liang and Altman, Ehud},
  journal = {Phys. Rev. Lett.},
  volume = {125},
  issue = {3},
  pages = {030505},
  numpages = {6},
  year = {2020},
  month = {Jul},
  publisher = {American Physical Society},
  doi = {10.1103/PhysRevLett.125.030505},
  url = {https://link.aps.org/doi/10.1103/PhysRevLett.125.030505}
}

@article{trans2,
  title = {Quantum Zeno effect and the many-body entanglement transition},
  author = {Li, Yaodong and Chen, Xiao and Fisher, Matthew P. A.},
  journal = {Phys. Rev. B},
  volume = {98},
  issue = {20},
  pages = {205136},
  numpages = {9},
  year = {2018},
  month = {Nov},
  publisher = {American Physical Society},
  doi = {10.1103/PhysRevB.98.205136},
  url = {https://link.aps.org/doi/10.1103/PhysRevB.98.205136}
}

@article{trans3,
  title = {Measurement-driven entanglement transition in hybrid quantum circuits},
  author = {Li, Yaodong and Chen, Xiao and Fisher, Matthew P. A.},
  journal = {Phys. Rev. B},
  volume = {100},
  issue = {13},
  pages = {134306},
  numpages = {26},
  year = {2019},
  month = {Oct},
  publisher = {American Physical Society},
  doi = {10.1103/PhysRevB.100.134306},
  url = {https://link.aps.org/doi/10.1103/PhysRevB.100.134306}
}

@article{trans4,
  title = {Measurement-Induced Phase Transitions in the Dynamics of Entanglement},
  author = {Skinner, Brian and Ruhman, Jonathan and Nahum, Adam},
  journal = {Phys. Rev. X},
  volume = {9},
  issue = {3},
  pages = {031009},
  numpages = {21},
  year = {2019},
  month = {Jul},
  publisher = {American Physical Society},
  doi = {10.1103/PhysRevX.9.031009},
  url = {https://link.aps.org/doi/10.1103/PhysRevX.9.031009}
}

@article{measonly1,
  title = {Entanglement Phase Transitions in Measurement-Only Dynamics},
  author = {Ippoliti, Matteo and Gullans, Michael J. and Gopalakrishnan, Sarang and Huse, David A. and Khemani, Vedika},
  journal = {Phys. Rev. X},
  volume = {11},
  issue = {1},
  pages = {011030},
  numpages = {23},
  year = {2021},
  month = {Feb},
  publisher = {American Physical Society},
  doi = {10.1103/PhysRevX.11.011030},
  url = {https://link.aps.org/doi/10.1103/PhysRevX.11.011030}
}

@article{measonly2,
  title = {Measurement-induced topological entanglement transitions in symmetric random quantum circuits},
  author = {Lavasani, Ali and Alavirad, Yahya and Barkeshli, Maissam},
  journal = {Nature Physics},
  volume = {17},
  issue = {3},
  pages = {342-347},
  numpages = {6},
  year = {2021},
  month = {Jan},
  publisher = {Nature},
  doi = {10.1038/s41567-020-01112-z},
  url = {https://www.nature.com/articles/s41567-020-01112-z}
}

@ARTICLE{measonly3,
       author = {{Jian}, Chao-Ming and {You}, Yi-Zhuang and {Vasseur}, Romain and {Ludwig}, Andreas W.~W.},
        title = "{Measurement-induced criticality in random quantum circuits}",
      journal = {\prb},
         year = 2020,
        month = mar,
       volume = {101},
       number = {10},
          eid = {104302},
        pages = {104302},
          doi = {10.1103/PhysRevB.101.104302},
archivePrefix = {arXiv},
       eprint = {1908.08051},
 primaryClass = {cond-mat.stat-mech},
       adsurl = {https://ui.adsabs.harvard.edu/abs/2020PhRvB.101j4302J}
}

@article{measonly4,
  title = {Measurement-protected quantum phases},
  author = {Sang, Shengqi and Hsieh, Timothy H.},
  journal = {Phys. Rev. Research},
  volume = {3},
  issue = {2},
  pages = {023200},
  numpages = {9},
  year = {2021},
  month = {Jun},
  publisher = {American Physical Society},
  doi = {10.1103/PhysRevResearch.3.023200},
  url = {https://link.aps.org/doi/10.1103/PhysRevResearch.3.023200}
}

@article{measonly5,
  title = {Measurement-induced criticality in $(2+1)$-dimensional hybrid quantum circuits},
  author = {Turkeshi, Xhek and Fazio, Rosario and Dalmonte, Marcello},
  journal = {Phys. Rev. B},
  volume = {102},
  issue = {1},
  pages = {014315},
  numpages = {9},
  year = {2020},
  month = {Jul},
  publisher = {American Physical Society},
  doi = {10.1103/PhysRevB.102.014315},
  url = {https://link.aps.org/doi/10.1103/PhysRevB.102.014315}
}

@ARTICLE{measonly6,
       author = {{Sierant}, Piotr and {Schir{\`o}}, Marco and {Lewenstein}, Maciej and {Turkeshi}, Xhek},
        title = "{Measurement-induced phase transitions in (d +1 ) -dimensional stabilizer circuits}",
      journal = {\prb},
         year = 2022,
        month = dec,
       volume = {106},
       number = {21},
          eid = {214316},
        pages = {214316},
          doi = {10.1103/PhysRevB.106.214316},
archivePrefix = {arXiv},
       eprint = {2210.11957},
 primaryClass = {cond-mat.stat-mech},
       adsurl = {https://ui.adsabs.harvard.edu/abs/2022PhRvB.106u4316S}
}

@article{measonly7,
  title = {Monitored quantum dynamics and the Kitaev spin liquid},
  author = {Lavasani, Ali and Luo, Zhu-Xi and Vijay, Sagar},
  journal = {Phys. Rev. B},
  volume = {108},
  issue = {11},
  pages = {115135},
  numpages = {26},
  year = {2023},
  month = {Sep},
  publisher = {American Physical Society},
  doi = {10.1103/PhysRevB.108.115135},
  url = {https://link.aps.org/doi/10.1103/PhysRevB.108.115135}
}

@article{measonly8,
  title = {Subsystem symmetry, spin-glass order, and criticality from random measurements in a two-dimensional Bacon-Shor circuit},
  author = {Sharma, Vaibhav and Jian, Chao-Ming and Mueller, Erich J.},
  journal = {Phys. Rev. B},
  volume = {108},
  issue = {2},
  pages = {024205},
  numpages = {10},
  year = {2023},
  month = {Jul},
  publisher = {American Physical Society},
  doi = {10.1103/PhysRevB.108.024205},
  url = {https://link.aps.org/doi/10.1103/PhysRevB.108.024205}
}

@article{tripartite,
  title = {Long-range multipartite entanglement near measurement-induced transitions},
  author = {Avakian, Sebastien J. and Pereg-Barnea, T. and Witczak-Krempa, William},
  journal = {Phys. Rev. Res.},
  volume = {7},
  issue = {2},
  pages = {023135},
  numpages = {10},
  year = {2025},
  month = {May},
  publisher = {American Physical Society},
  doi = {10.1103/PhysRevResearch.7.023135},
  url = {https://link.aps.org/doi/10.1103/PhysRevResearch.7.023135}
}

@misc{tripartite1,
  doi = {10.48550/ARXIV.2407.03206},
  url = {https://arxiv.org/abs/2407.03206},
  author = {Xu,  Guanglei and Zhang,  Yu-Xiang},
  title = {Multipartite Greenberger-Horne-Zeilinger Entanglement in Monitored Random Clifford Circuits},
  publisher = {arXiv},
  year = {2024},
  copyright = {Creative Commons Attribution 4.0 International}
}

@article{qfimonitored,
  title = {Multipartite Entanglement Structure of Monitored Quantum Circuits},
  author = {Lira-Solanilla, Arnau and Turkeshi, Xhek and Pappalardi, Silvia},
  journal = {Phys. Rev. Lett.},
  volume = {135},
  issue = {8},
  pages = {080401},
  numpages = {9},
  year = {2025},
  month = {Aug},
  publisher = {American Physical Society},
  doi = {10.1103/fl34-h1p1},
  url = {https://link.aps.org/doi/10.1103/fl34-h1p1}
}

@article{qfimonitored1,
  title = {Multipartite entanglement in the measurement-induced phase transition of the quantum Ising chain},
  author = {Paviglianiti, Alessio and Silva, Alessandro},
  journal = {Phys. Rev. B},
  volume = {108},
  issue = {18},
  pages = {184302},
  numpages = {7},
  year = {2023},
  month = {Nov},
  publisher = {American Physical Society},
  doi = {10.1103/PhysRevB.108.184302},
  url = {https://link.aps.org/doi/10.1103/PhysRevB.108.184302}
}

@article{qfimonitored2,
  title = {Metrology and multipartite entanglement in measurement-induced phase transition},
  volume = {8},
  ISSN = {2521-327X},
  url = {http://dx.doi.org/10.22331/q-2024-04-30-1326},
  DOI = {10.22331/q-2024-04-30-1326},
  journal = {Quantum},
  publisher = {Verein zur Forderung des Open Access Publizierens in den Quantenwissenschaften},
  author = {Di Fresco,  Giovanni and Spagnolo,  Bernardo and Valenti,  Davide and Carollo,  Angelo},
  year = {2024},
  month = apr,
  pages = {1326}
}

@article{qfimonitored3,
  title = {Measurement-induced multipartite-entanglement regimes in collective spin systems},
  volume = {8},
  ISSN = {2521-327X},
  url = {http://dx.doi.org/10.22331/q-2024-01-18-1229},
  DOI = {10.22331/q-2024-01-18-1229},
  journal = {Quantum},
  publisher = {Verein zur Forderung des Open Access Publizierens in den Quantenwissenschaften},
  author = {Poggi,  Pablo M. and Muñoz-Arias,  Manuel H.},
  year = {2024},
  month = jan,
  pages = {1229}
}

@misc{kparty,
      title={Spatial structure of multipartite entanglement at measurement induced phase transitions}, 
      author={James Allen and William Witczak-Krempa},
      year={2025},
      eprint={2509.12109},
      archivePrefix={arXiv},
      primaryClass={quant-ph},
      url={https://arxiv.org/abs/2509.12109}, 
}

@article{qfi,
  title = {Fisher information and multiparticle entanglement},
  author = {Hyllus, Philipp and Laskowski, Wies\l{}aw and Krischek, Roland and Schwemmer, Christian and Wieczorek, Witlef and Weinfurter, Harald and Pezz\'e, Luca and Smerzi, Augusto},
  journal = {Phys. Rev. A},
  volume = {85},
  issue = {2},
  pages = {022321},
  numpages = {10},
  year = {2012},
  month = {Feb},
  publisher = {American Physical Society},
  doi = {10.1103/PhysRevA.85.022321},
  url = {https://link.aps.org/doi/10.1103/PhysRevA.85.022321}
}

@misc{nonlocalqfi,
      title={Local and Non-local Entanglement Witnesses of Fermi Liquid}, 
      author={Yiming Wang and Yuan Fang and Fang Xie and Qimiao Si},
      year={2025},
      eprint={2502.13958},
      archivePrefix={arXiv},
      primaryClass={cond-mat.str-el},
      url={https://arxiv.org/abs/2502.13958}, 
}

@article{entstructures,
  title = {Multipartite entanglement structures in quantum stabilizer states},
  author = {Sharma, Vaibhav and Mueller, Erich J.},
  journal = {Phys. Rev. A},
  volume = {112},
  issue = {1},
  pages = {012411},
  numpages = {10},
  year = {2025},
  month = {Jul},
  publisher = {American Physical Society},
  doi = {10.1103/1cqt-8rxf},
  url = {https://link.aps.org/doi/10.1103/1cqt-8rxf}
}

@inbook{theil,
  title = {A Rank-Invariant Method of Linear and Polynomial Regression Analysis},
  ISBN = {9789401125468},
  ISSN = {1570-5811},
  url = {http://dx.doi.org/10.1007/978-94-011-2546-8_20},
  DOI = {10.1007/978-94-011-2546-8_20},
  booktitle = {Henri Theil’s Contributions to Economics and Econometrics},
  publisher = {Springer Netherlands},
  author = {Theil,  Henri},
  year = {1992},
  pages = {345–381}
}

@article{sen,
  title = {Estimates of the Regression Coefficient Based on Kendall’s Tau},
  volume = {63},
  ISSN = {1537-274X},
  url = {http://dx.doi.org/10.1080/01621459.1968.10480934},
  DOI = {10.1080/01621459.1968.10480934},
  number = {324},
  journal = {Journal of the American Statistical Association},
  publisher = {Informa UK Limited},
  author = {Sen,  Pranab Kumar},
  year = {1968},
  month = dec,
  pages = {1379–1389}
}

@article{gottknill,
  title = {Improved simulation of stabilizer circuits},
  author = {Aaronson, Scott and Gottesman, Daniel},
  journal = {Phys. Rev. A},
  volume = {70},
  issue = {5},
  pages = {052328},
  numpages = {14},
  year = {2004},
  month = {Nov},
  publisher = {American Physical Society},
  doi = {10.1103/PhysRevA.70.052328},
  url = {https://link.aps.org/doi/10.1103/PhysRevA.70.052328}
}

@article{cantor,
  title = {Void distribution of random Cantor sets},
  author = {Tremblay, R. R. and Siebesma, A. P.},
  journal = {Phys. Rev. A},
  volume = {40},
  issue = {9},
  pages = {5377--5381},
  numpages = {0},
  year = {1989},
  month = {Nov},
  publisher = {American Physical Society},
  doi = {10.1103/PhysRevA.40.5377},
  url = {https://link.aps.org/doi/10.1103/PhysRevA.40.5377}
}

@article{thielsenadv,
  title = {The Theil-Sen Estimator with Doubly Censored Data and Applications to Astronomy},
  volume = {90},
  ISSN = {1537-274X},
  url = {http://dx.doi.org/10.1080/01621459.1995.10476499},
  DOI = {10.1080/01621459.1995.10476499},
  number = {429},
  journal = {Journal of the American Statistical Association},
  publisher = {Informa UK Limited},
  author = {Akritas,  Michael G. and Murphy,  Susan A. and Lavalley,  Michael P.},
  year = {1995},
  month = mar,
  pages = {170–177}
}

@article{mad,
  title = {Detecting outliers: Do not use standard deviation around the mean,  use absolute deviation around the median},
  volume = {49},
  ISSN = {0022-1031},
  url = {http://dx.doi.org/10.1016/j.jesp.2013.03.013},
  DOI = {10.1016/j.jesp.2013.03.013},
  number = {4},
  journal = {Journal of Experimental Social Psychology},
  publisher = {Elsevier BV},
  author = {Leys,  Christophe and Ley,  Christophe and Klein,  Olivier and Bernard,  Philippe and Licata,  Laurent},
  year = {2013},
  month = jul,
  pages = {764–766}
}

@article{fragment1,
  title = {Coagulation and fragmentation: Universal steady‐state particle‐size distribution},
  volume = {42},
  ISSN = {1547-5905},
  url = {http://dx.doi.org/10.1002/AIC.690420612},
  DOI = {10.1002/aic.690420612},
  number = {6},
  journal = {AIChE Journal},
  publisher = {Wiley},
  author = {Spicer,  Patrick T. and Pratsinis,  Sotiris E.},
  year = {1996},
  month = jun,
  pages = {1612–1620}
}

@article{fragment2,
  title = {Particle-based modeling of aggregation and fragmentation processes: Fractal-like aggregates},
  volume = {240},
  ISSN = {0167-2789},
  url = {http://dx.doi.org/10.1016/j.physd.2011.01.003},
  DOI = {10.1016/j.physd.2011.01.003},
  number = {9–10},
  journal = {Physica D: Nonlinear Phenomena},
  publisher = {Elsevier BV},
  author = {Zahnow,  Jens C. and Maerz,  Joeran and Feudel,  Ulrike},
  year = {2011},
  month = apr,
  pages = {882–893}
}

@misc{dataset,
  doi = {10.5281/ZENODO.20147205},
  url = {https://zenodo.org/doi/10.5281/zenodo.20147205},
  author = {Sharma,  Vaibhav},
  title = {Entanglement Depth dataset},
  publisher = {Zenodo},
  year = {2026},
  copyright = {Creative Commons Attribution 4.0 International}
}

\end{document}